\documentclass[twocolumn,showpacs,preprintnumbers,amsmath,amssymb]{revtex4}

\newcommand{\be}{\begin{equation}}
\newcommand{\ee}{\end{equation}}
\newcommand{\bea}{\begin{eqnarray}}
\newcommand{\eea}{\end{eqnarray}}

\newcommand{\p}{\partial}

\usepackage{graphicx}
\usepackage{dcolumn}
\usepackage{bm}

\begin{document}

\preprint{}

\title{Yang Mills Magneto-Fluid Unification}
\author{Bindu A. Bambah}
\affiliation{School of Physics, \\University of Hyderabad, \\
Hyderabad,  Andhra Pradesh,  500 046, India }
\author{Swadesh M. Mahajan}
\affiliation{Institute of Fusion Studies,\\ University of Texas,
\\ Austin, Texas, 78712 U.S.A \\ and }
\author{Chandrasekher Mukku}
\affiliation{International Institute of Information Technology,\\
Hyderabad, Andhra Pradesh,  500 032\\ India}


\begin{abstract}
We generalize the hybrid magneto-fluid model of a charged fluid
interacting with an electromagnetic field to the dynamics of a
relativistic hot fluid interacting with a non-Abelian field. The
fluid itself is endowed with a non-Abelian charge and the
consequences of this generalization are worked out. Applications of
this formalism to the Quark Gluon Plasma are suggested.
\end{abstract}



\maketitle

\section{Introduction}

Recent experiments at the relativistic heavy ion collider (RHIC)
have shed light on the behavior of hot dense nuclear matter
\cite{expt}. The conjecture,  that quarks and gluons de-confine and
become a plasma in extreme conditions \cite{asym},  is close to
being experimentally proved. However,  it has been realized that the
de-confined quark gluon matter that has been revealed at RHIC,  far
from being the weakly interacting collisionless plasma envisioned by
theorists,  is,  in fact,  behaving more like a quark gluon liquid,
or a strongly interacting plasma \cite{shuryak}. The QGP liquid (or
strongly interacting plasma) is dense,  but seems to flow with very
little viscosity. It flows so freely that it approximates an ideal,
or perfect fluid,  the kind governed by the standard laws of
hydrodynamics. Thus,  most of the phenomenological input for the
explanation of the data at RHIC comes from adopting a hydrodynamic
approach to the plasma and consequently,  fluid dynamics is used to
deduce its properties. For  a proper fluid dynamical description of
the quark gluon fluid the fact that quarks and gluons have a
non-Abelian charge  has to be taken into consideration. There have
been theories focusing on various aspects of an ideal fluid  in
interaction with Yang-Mills fields called Yang Mills
magnetohydrodynamics or chromohydrodynamics \cite{Heinz,  Gyulassy,
Choquet-Bruhat, Holm,  Van Hutten, blaizot}. The relativistic heavy
ion experiments coupled with recent studies by Jackiw and his
collaborators on the "Clebsch representation" of Yang Mills fluid
dynamics have lead to a resurgence of interest in these theories
\cite{expt, jackiw1, jackiw2}. In tune with this revival,  we
investigate the dynamics of a relativistic hot fluid with a
non-abelian charge in terms of a model which unifies the Yang Mills
field with the flow field strength tensor \cite{mahajan1}. Apart
from its possible phenomenological applications,  the motivation for
the following work is based on the aesthetic criterion of unifying
the fluid field and the Yang Mills field into a Yang Mills
"magneto-fluid" by a "gauge principle". The similarities between one
gauge field theory - electromagnetism and fluid dynamics have been
explored extensively and fluid flow has been shown to have a formal
equivalence with a gauge theory \cite{mahajan1, indo}.
 It has also been shown that  a fully
antisymmetric flow tensor, resembling the electromagnetic field, can
be constructed and the unification is achieved by  defining an
effective field strength tensor that combines appropriately weighted
electromagnetic and flow fields  \cite{mahajan1}. Is a consistent
and useful non-Abelian generalization of this genre of flow-field
unification possible? This investigation constitutes the theme of
this paper.

\section{Abelian Magneto-fluid Unification.}

First,  let us recapitulate the salient features of Abelian(Maxwell)
Magneto-fluid unification \cite{mahajan1}. Although Maxwells
electrodynamics provides equations of motion for the electric and
magnetic fields, for describing their interaction with matter fields
(charged particles), the Lorentz force law has to be independently
postulated. In contrast, in a gravity coupled plasma, a natural
consequence of general covariance is the conservation of energy and
momentum,  and the Lorentz force law for charged particles moving in
a gravitational field \cite{MTW} can be derived from the field
equations to be :
\begin{equation}
\nabla_{\bf{U}}\bf{U}=\frac{q}{m}\bf{U}\cdot F, \end{equation}
where,  $\bf{U}$ is the tangent vector to a geodesic (the velocity
vector). In the limit of a weak gravitational field (the flat
space-time limit,  $\nabla_{\mu}\rightarrow\partial_{\mu}$  ), the
component form of Eq.(1) reads :
\begin{equation}U^{\mu}\partial_{\mu}U^{\nu}=\frac{q}{m}U^{\mu}F^{\nu}{}_{\mu}\end{equation}
Because of  the antisymmetry of $F_{\mu\nu}$, contraction  with
$U^{\nu}$  reduces the right hand side of Eq.(2) to zero. Thus, with
no loss of generality, the coefficient of $U^{\mu}$,  on the left
hand side of the equation, should also be anti-symmetrized. Since
$U^{\mu}$, the four-velocity, obeys $U^{\mu}U_{\mu}=-1$ (implying
$U^{\mu}\partial_{\nu}U^{\mu}=0$),
 Eq.(2) may be written as \begin{equation}
U^{\mu}\partial_{\mu}U^{\nu}-U^{\mu}\partial^{\nu}U_{\mu}=\frac{q}{m}U^{\mu}F^{\nu}{}_{\mu}\label{lorentz}\end{equation}
With the definition
\begin{equation}P_{\mu\nu}=\partial_{\mu}U_{\nu}-\partial_{\nu}U_{\mu}.\end{equation}
we can write Eq.(3) as
\begin{equation}U^{\mu}(\frac{m}{q}P_{\mu\nu}+F_{\mu\nu})=0.\end{equation}
For deriving equations of motion for point
particles,  usually a limiting procedure is invoked \cite{MTW}.
For the motion of fluids, however,
 a small volume element of the fluid is the limiting element and
the statistical properties of the fluid come into play. Recently
\cite{mahajan1} suggested that the $S_{\mu\nu}=\partial_{\mu}
fU_{\nu}-\partial_{\nu}fU_{\mu}$ must replace the particle
$P_{\mu\nu}$ for a new and natural "minimal coupling" to describe a
fluid interacting with the Maxwell fields,where $f$ represents a
temperature dependent statistical attribute of the fluid, and is
related to the enthalpy $h$, number density $n$ and mass $m$ of the
fluid by the relation $ h=mnf(T).$  In terms of $S_{\mu\nu}$, the
"fluid Lorentz equation" derived in \cite{mahajan1} is
 \be
T\p^{\nu}\sigma=q(F^{\mu\nu}+\frac{m}{q}S^{\mu\nu})U_{\mu}
 \ee
 where, $\sigma$ is an entropy density.
The limiting procedure to the point particle case, then, is simply
equivalent to $fU^{\mu}\rightarrow U^{\mu}$, $S_{\mu\nu}\rightarrow
P_{\mu\nu}$, as $f\longrightarrow 1$ ($T\longrightarrow 0$).

 The curvature
$F_{\mu\nu}$ corresponding to the connection $A_{\mu}$  is obtained
from the commutator of two covariant derivatives
$D_{\mu}=\partial_{\mu}-iqA_{\mu}$. In a similar vein,  we can
define a ``unified'' connection
$Q^{\mu}=A^{\mu}+\frac{m}{q}fU^{\mu}$, which corresponds to a
minimally coupled  hot magnetofluid. The new Abelian covariant
derivative for the unified field
\begin{equation}
D_{\mu}=\partial_{\mu}-iqQ_{\mu}
=\partial_{\mu}-iqA_{\mu}-imfU_{\mu}.\end{equation}
leads to the unified curvature
\begin{equation}[D_{\mu}, D_{\nu}]=-iqF_{\mu\nu}-imS_{\mu\nu}, \end{equation}
where  $F_{\mu\nu}=\partial_{\mu}A_{\nu}-\partial_{\nu}A_{\mu}$ and
$S_{\mu\nu}=\partial_{\mu} fU_{\nu}-\partial_{\nu}fU_{\mu}$

This new,  temperature dependent  minimal coupling procedure has
many interesting consequences which are explored in \cite{mahajan1,
mahajan2}.

\section{The Yang Mills Magneto-Fluid Tensor}

Now we return to the main theme of this work-the dynamics of  a non-Abelian fluid. The
 non-Abelian gauge field is represented by
\begin{equation}F^{a}{}_{\mu\nu}=\partial_{\mu}A^{a}{}_{\nu}-\partial_{\nu}A^{a}{}_{\mu}-ig[A_{\mu}, A_{\nu}]^{a}.\end{equation}
where, $[A_{\mu},
A_{\nu}]^{a}=iC^{a}_{bc}A^{b}{}_{\mu}A^{c}{}_{\nu}$,  $C^{a}{}_{bc}$
are the structure constants of the gauge group and $g$ is the gauge
"charge".

Since $F$ now has a gauge index,  the right hand side of the Abelian
equations of motion suggests a generalization of the fluid flow
vector to include a gauge,  or a non-Abelian index. The RHS of the
equations (\ref{lorentz}) can then be written as : $
 (q/m)U^{\mu}{}_{a}F^{a}{}^{\nu}{}_{\mu}$. Correspondingly,
the LHS,  $U^{\mu}\partial_{\mu}U^{\nu}$ requires a non-Abelian
generalization. This mandates giving the flow field $U_{\mu}$ a
non-Abelian index,  and we are led to a generalization of the
Abelian flow tensor $S_{\mu \nu}$ to $S^{a}{}_{\mu\nu}$.

 Following the Abelian route,  an explicit form of  $S^{a}{}_{\mu\nu}$  is given
by evaluating the non-Abelian curvature for the generalized
non-Abelian covariant derivative,
\begin{equation}D_{\mu}=\partial_{\mu}-ig[A_{\mu},  ]-im[fU_{\mu},
].\label{totcovder}\end{equation} The corresponding curvature
 \begin{equation}[D_{\mu}, D_{\nu}]^{a}=-igF^{a}_{\mu\nu}-imS^{a}_{\mu\nu}\end{equation}
coupled with Eq.9, defines the YM generalization of  $S_{\mu \nu}$
\begin{eqnarray}S^{a}_{\mu\nu}&=&\partial_{\mu}fU^{a}{}_{\nu}-\partial_{\nu}fU^{a}{}_{\mu}-imf^{2}[U_{\mu}, U_{\nu}]^{a} \nonumber \\&-&igf[A_{\mu}, U_{\nu}]^{a}-igf[U_{\mu}, A_{\nu}]^{a}.\label{nas}\end{eqnarray}
written, succintly,  as
\begin{equation}
S^{\mu\nu}_{ a}= {\cal{D}}^{\mu}(fU^{\nu}_
{a})-{\cal{D}}^{\nu}(fU^{\mu}_{ a})-imf^2[U^{\mu}_{ b}, U^{\nu}_{
c}]
\end{equation} where ${\cal{D_{\mu}}}=\partial_{\mu}-ig[A_{\mu}, ~]$
is the ordinary non-Abelian gauge covariant derivative.  Notice that $S^{\mu\nu}_{ a}$
encompasses the pure flow-field as well as the interaction.

In the presence of a matter gauge current, the Yang Mills Field  evolves as
\begin{equation}\label{YM 1}
  {\cal{D}}_{\mu}F_{a}^{\mu\nu}  =
  -J^{\nu}_a.
  \end{equation}

The L.H.S of Eq.(14) is easily related to the energy momentum tensor of the
 gauge field  \be
\Theta_{\mu \nu}=F^{\mu}_{a\sigma}F^{\nu
\sigma}_{a}-\frac{1}{4}\eta^{\mu\nu}F^{\rho \sigma}_{a}F_{\rho
\sigma a} \ee through the Bianchi identity \be
{\cal{D}}^{\nu}F^{\sigma \rho}+{\cal{D}}^{\sigma}F^{
\rho\nu}+{\cal{D}}^{\rho}F^{\nu \sigma}=0,  \ee  leading to \be
\p^{\nu}\Theta_{\nu\mu}=-F^{a}_{\nu\mu}{\cal{D}}_{\rho}F_{a}^{\rho\nu}
=F^{a}{}_{\mu\nu} J^{\nu}_a.\label{feq}
 \ee
The evolution of  Yang Mills  potential $A_{\mu}=A^{a}_{\mu} T_a$ is
dictated by the matter current $J_{\mu}$ in addition to the  inherent nonlinearities in field equations.

 In the conventional particle description,  the current $J_{\mu}$ is
 constructed from the wave function,  which itself evolves
 covariantly in the gauge field.
 Ideally,  for  strongly interacting matter such as a quark gluon
 plasma,  such  a current has to be constructed from a
 collective  many-body wave function.
 This procedure is very cumbersome and sometimes
not very illuminating. Thus,  a fluid description,  in terms of flow
and thermodynamic variables,  is useful,  because, it captures a
complicated dynamics in terms of a few collective
variables.   We are "viewing" a strongly interacting
 many particle system through a set of representative "flow"
 fields. We consider not just one but
 several flow fields (labeled by a species index ${s}$) denoting  the
 particles (quarks,  anti-quarks,  gluons)  interacting with the non-Abelian gauge field.
 Each species can in principle have different charges,
densities,  temperatures etc. Each species is taken to be a perfect
fluid with an energy momentum tensor of the form
\begin{equation}
T^{\mu\nu}_{{s}}=p_{{s}}\eta^{\mu\nu}+h_{{s}}U_{a, s}^{\mu}U^{\nu a,
s},
\end{equation}
where,  $p_s$ is the pressure, and the enthalpy density  $h_s$ is given by
\begin{equation}
h_s=m_s n_{R,s}\frac{K_{3}(m_s/T)}{K_{2}(m_s/T)}=m_s n_{R, s}f_s(T)
\label{ent},
\end{equation} and $n_{R,s}$ measures the density in the rest frame for the given species.

Each of the non-Abelian species contributes a  flux
\begin{equation}
\Gamma^{\mu}{}_{a, s}=n_{R, s}U_{a, s}^{\mu}\label{gamma},
\end{equation}
towards  the total non-Abelian current \be J_{a}^{\mu}=\sum_{s}
g_s\Gamma^{\mu}{}_{a, s}\label{curr},\ee  where $g_s$ is the gauge
coupling for the species `s". The matter fields contributing to
$J^{\mu}_{a}$
 evolve covariantly under the action of the
 non-Abelian covariant derivative. The multi-species formulation allows the possibility
 $\sum_{s}g_{s}n_{R, s}=0$, that is,  the charge density can be zero in the
 rest frame  of the fluid.

In Eq.21, we have already identified the charge weighted flux sum
(over matter species) with the current $J_{a}^{\mu}$ that drives the gauge field equation(14).
This identification is perfectly sensible  and is consistent with
 the  choice of the fluid  equation of motion
\begin{equation}\label{fluid 1}
\partial_{\mu}T^{\mu\nu}_{a}= g_s n_{R, s}U^{\nu}{}_{a, {s}}.
\end{equation}
Summing Eq.22 over species, we obtain
\begin{equation}
\sum_{s}\partial_{\mu}T^{\mu\nu}_{s}=\p_{\mu}{\cal{T^{\mu\nu}}}=\sum_{s}
g_{s}F_a^{\nu\sigma}\Gamma_{\sigma}^{a,
s}=F_{a}^{\nu\sigma}J_{\sigma}^{a}\label{curcont},\end{equation}
where ${\cal{T^{\mu\nu}}}$ is the total fluid tensor.  Combining it with (\ref{feq}),
we arrive at  the expected conservation law for the total energy momentum
tensor ( matter plus field)
\begin{equation}
\p^{\mu}(\Theta_{\mu\nu}+{\cal{T_{\mu\nu}}})=0\label{totcovder},
\end{equation}
justifying the expression for the current.
For the rest of the paper,  we shall drop the  species index unless it is essential for clarity.

From the continuity equation,  generalized to the non-Abelian case,
\begin{equation}{\cal{D}}_{\mu}^{a}\Gamma^{\mu}{}_a=0. \end{equation}
 or
\begin{equation}\partial_{\mu}(n_{R}U^{\mu}{}^{a})=-gn_{R}C^{abc}A^{b}{}_{\mu}U^{c\mu}.\end{equation}
and from the the definition of $T_{\mu\nu}$  (for a perfect fluid),
we have
\begin{eqnarray}
\partial_{\mu}T^{\mu\nu}&=&\partial^{\nu}p+m\partial_{\mu}(fn_{R}U_{a}{}^{\mu}U_{a}{}^{\nu})\nonumber\\
                        &=&\partial^{\nu}p+mn_{R}U_{a}{}^{\mu}\partial_{\mu}(fU_{a}{}^{\nu})
                        -gn_{R}mfU_{a}{}^{\nu}A_{b\mu}U^{b\mu}C^{abc}\nonumber\\
                        &=&\partial^{\nu}p
                        +mn_{R}U_{a\mu}[{\cal{D}}^{\mu}fU^{\nu}]_{a}\label{conT}
\end{eqnarray}
The second term on the R.H.S of Eq.27 is related to   $S_{a}{}^{\mu\nu}$ given by equation(\ref{nas}):
 \begin{equation}
U_{a\mu}S_{a}{}^{\mu\nu}=N\partial^{\nu}f
+U_{a\mu}[{\cal{D}}^{\mu}fU^{\nu}]_{a}\label{uconts}
\end{equation}
Combining equations(\ref{conT}) and (\ref{uconts}),
\begin{equation}
\partial_{\mu}T^{\mu\nu}=\partial^{\nu}p+mn_{R}(U_{a\mu}S_{a}{}^{\mu\nu}-N\partial^{\nu}f)
\end{equation}
where $N$ stands for the dimension of the gauge group and we have
applied $U^{a\mu}U_{a\mu}=-N$ as a natural generalization of
$U^{\mu}U_{\mu}=-1$ in the Abelian case. Substituting appropriately
from equations (\ref{gamma},\ref{curr}) and (\ref{curcont}),  we
find
\begin{equation}
\partial^{\nu}p-Nmn_{R}\partial^{\nu}f=gn_{R}[F_{a}{}^{\mu\nu}+\frac{m}{g}S_{a}{}^{\mu\nu}]U_{a\mu}\label{eom}
\end{equation}

In analogy with the Abelian case \cite{mahajan1}, we define the new
unified non-Abelian tensor \be M^{\mu \nu}_{a}=F^{\mu
\nu}_{a}+\frac{m}{g}S^{\mu \nu}_{a},  \ee representing the matter and
 gauge field (including their interaction). Then,
 equation (\ref{eom}) becomes
\begin{equation}
\partial^{\nu}p-Nmn_{R}\partial^{\nu}f=gn_{R}M^{\mu\nu}_{a}U_{a\mu}\label{eom1}
\end{equation}
 Defining the entropy $\sigma$ from Eqn. (\ref{ent}), in analogy with \cite{mahajan1},
$\sigma=ln[(p/K_2)(\frac{m}{T})^2exp(-\frac{mK_3}{TK_2})],$
 Eqs.30-32 may be combined to yield
\begin{equation}
T\p^{\nu}\sigma=gM^{\mu\nu}_{a}U_{a\mu}\label{eom3}
\end{equation}
 For a homentropic fluid (a relevant limit for the QGP),  the equation of motion becomes even simpler:
 \begin{equation}
 gM^{\mu \nu}{}_a U_{\nu}{}_{a}=
[F_{a}{}^{\mu\nu}+\frac{m}{g}S_{a}{}^{\mu\nu}]U_{a\mu}=0\label{eom2}
\end{equation}

We have just shown the existence of  a unified "mininimally" coupled potential for
hot non-Abelian fluids \be
Q^{\mu}_{a}=A^{\mu}_a+\frac{m}{g}fU^{\mu}, \ee
 with its corresponding field tensor
 \be
 M_{a}^{\mu\nu}=\p_{\mu} Q_a^{\nu}-\p_{\nu}
 Q^{\mu}_a+gc^{bc}_{a}Q^{\mu}_bQ^{\nu}_c.
 \ee
 Through $M_a^{\mu\nu}$ and $Q^{\mu}_a$,  we have put the
 non-Abelian flow field and the non-Abelian gauge field on the
 same footing. The unification opens up an opportunity to apply the powerful
 machinery of gauge theories to the unified gauge- flow field; this formulation
complements the previous work on the subject \cite{ Holm, jackiw2}.

 The equation of motion  given by
(\ref{eom2}) is the analogue of the two equations that Jackiw
\cite{jackiw2} finds for the continuum version of Wong's
equations. However,  unlike Jackiw's generalization, which seems to
couple Yang-Mills fields with a fluid of non-Abelian Charges,  our
proposal provides a natural non-linearity within the coupled
system since $S^{a}{}_{\mu\nu}$ contains an interaction with
the Yang-Mills fields through the connection $A^{a}{}_{\mu}$. This
natural non-linearity provides  a mechanism to bypass the
theorem forbidding the existence of solitonic configurations in a
plasma \cite{freidberg}.

\section{Topological Invariants of the Yang-Mills "Magneto-Fluid".}

In order to explore any interesting
consequences of this formalism for Yang-Mills fluids,
let us return first to the Abelian formalism and
results on helicity conservation.

The spatial components of the equation of motion in the Abelian case
may be spelled out as:
\begin{equation}U_{0}(\frac{m}{q}S^{0i}+F^{0i})+U_{j}(\frac{m}{q}S^{ji}+F^{ji})=0 \label{mag}\end{equation}
Since $F^{0i}$ is just the electric field,  let us call the
combined factor $(\frac{m}{q}S^{0i}+F^{0i})$,  the
fluid-generalized electric field $\hat{E^{i}}$. Let us make the
same prescription for the magnetic components. Then, since $U_{0}$
is just the relativistic factor,  $\gamma$,  and
$U_{i}=\gamma\vec{U}$,  eq. (\ref{mag}) corresponds to:
\begin{equation}\gamma
\hat{\vec{E}}+\gamma\vec{U}\times\hat{\vec{B}}=0.\end{equation} An
immediate consequence is the condition
\begin{equation}\hat{\vec{E}}\cdot\hat{\vec{B}}=0\label{edotb}.\end{equation}
Since the product of $M_{\mu\nu}=\frac{m}{q}S_{\mu\nu}+F_{\mu\nu}$
with its dual
${\cal{M}}_{\mu\nu}=\frac{1}{2}\epsilon_{\mu\nu}{}_{\lambda\rho}M^{\lambda\rho}$,
is proportional to $\hat{\vec{E}}\cdot\hat{\vec{B}}$,
Eq.(\ref{edotb}) demands
\begin{equation}\frac{1}{2}M^{\mu\nu}{\cal{M}}_{\mu\nu}=0.\end{equation}
Since the left hand side of this equation is a boundary term,  we
can assume the existence of a $K^{\mu}$, which for electrodynamics,
is conserved
\begin{equation}\partial_{\mu}K^{\mu}=0.\end{equation}
 It is well known that $K^{\mu}$ gives
rise to helicity conservation and $K^{\mu}$ is identified with the
fluid field equivalent of the  Abelian Chern-Simons vector
$A_{\mu}{\cal{F}}^{\mu\nu}$ (see \cite{mahajan1}). The quantity
$\frac{1}{2}M^{\mu\nu}{\cal{M}}_{\mu\nu}$ represents the
topological winding number (charge) of solutions to the fluid
field equations of motion and the fact that it is zero  supports
the widely held view that an ordinary electron positron plasma,
for example, does not support stable, self confining knot like
solutions. This is upheld by  a virial theorem due to Shafranov
which states that  a static configuration of a plasma in isolation
is dissipative. Recently in \cite{fadeev} it has been proposed
that this no go theorem is circumvented by introducing non linear
interactions. We shall now show that the generalization to the YM
plasma overides this limitation and can support stable knot like
solutions.

For the non-Abelian case,  the spatial part of  equation(\ref{eom2})
is some what more complicated: \be
U_{0}{}^{a}(\frac{m}{q}S_{a}{}^{0i}+F_{a}{}^{0i})+U^{a}{}_{j}(\frac{m}{q}S_{a}{}^{ji}+F_{a}{}^{ji})=0\label{nabvec}\ee
To manipulate the Eq. (\ref{nabvec}), we have to face the question
of factoring the non-Abelian four-velocity $U_{\mu}{}^{a}$. At this
stage there is no immediate compulsion for a factoring out of the
generators (charges) of the gauge group. Instead,  one may assume
that $U^{a}{}_{\mu}U_{a}{}^{\mu}=tr_{group}U_{\mu}U^{\mu}=-N$ ($N$
being the dimension of the gauge group) implies the existence of a
full non-Abelian flow and that the velocity four-vector is
normalized in each flow. In terms of the fluid-generalized electric
and magnetic fields, equation(\ref{nabvec}) becomes: \be
\sum_{a}\gamma\hat{\vec{E_{a}}}+\gamma\vec{U^{a}}\times\hat{\vec{B}}_{a}=0\ee
Because of the trace over the group indices, Eq.(\ref{nabvec})
implies that unlike in the Abelian case, the product $
M^{a}_{\mu\nu}{\cal{M}}_{a}{}_{\mu\nu}\ne 0$. Consequently, the non
Abelian generalization of the topological charge is not necessarily
zero opening up the possibility of non trivial topological
structures being supported by the non-Abelian fluid plasma.

However, we may define a generalized Chern-Simons vector,
$C^{\mu}$ as  \be
\partial_{\mu}C^{\mu}=\frac{1}{2}M^{a}_{\lambda\rho}{\cal{M}}_{a}{}_{\lambda\rho}.\label{nabcs}\ee
In terms of the connection $ Q^{\mu}$, the generalized Chern-Simons
vector $C{\mu}$ is given by:\be
C^{\mu}=Q^{a}{}_{\nu}[{\cal{M}}^{a\mu\nu}-\frac{g}{6}\epsilon^{\mu\nu\lambda\rho}C^{abc}Q_{b\lambda}Q_{c\rho}].\ee

To analyze this topological term,  we  look at the gauge
properties of this potential $Q^{a}{}_{\mu}$. Under a gauge
transformation $\Omega_g$, \be
{\bf{A'_{\mu}}}=\Omega_{g}{\bf{A_{\mu}}}\Omega_g^{-1}-\frac{i}{g}(\partial_{\mu}\Omega_g)\Omega_g^{-1},
\ee it is not difficult to see that: \be
{\bf{Q'_{\mu}}}=\Omega_g{\bf{Q_{\mu}}}\Omega_g^{-1}-\frac{i}{g}(\partial_{\mu}\Omega_g)\Omega_g^{-1}.\ee
This means that the non-Abelian fluid velocity vector,
$U^{a}{}_{\mu}$ transforms covariantly \be
{\bf{U'_{\mu}}}=\Omega_g{\bf{U_{\mu}}}\Omega_g^{-1}\ee under a
gauge transformation and $Q^{a}{}_{\mu}$ can hence be strictly
identified with a non-Abelian gauge connection. An immediate
consequence for our analysis is that the $C_{\mu}$ that we have
written above will indeed give us a generalized Chern-Simons
invariant associated with $Q^{a}{}_{\mu}$. In addition,  the
Yang-Mills connection $A^{a}{}_{\mu}$ will provide us with the
standard Chern-Simons invariant.

The important point is that the minimal coupling procedure
introduced in \cite{mahajan1} and generalized here to the
non-Abelian case,  allows us to link these two Chern-Simons terms
through the fluid velocity vector.  Unlike the Abelian
case (where the divergence of the generalized helicity four vector for the combined system was
forced to be zero),  we have two
topological quantities  in the non-Abelian case:  one coming from the combined fluid+YM
case and one from the YM case with the fluid velocity vector tying
the two together.

To understand the topological implications of our result, consider
the transformation of the quantity $\frac{1}{8\pi}\int C_{\mu}d^3x$
under gauge transformations. For this we use the wedge product
notation $A\wedge F=\epsilon^{ijk}A_iF_{jk}d^3x$ to write \bea
I&=&\frac{1}{8\pi^2}\int C_{\mu}d^3x=\frac{1}{8\pi^2}\int Tr(Q\wedge
dQ-\frac{2}{3}Q\wedge Q\wedge Q)\nonumber \\&=&\frac{1}{8\pi^2}\int
Tr(Q\wedge M+\frac{1}{3}Q\wedge Q\wedge Q) \eea Using the
transformation properties established for Q we get the gauge
transformed $I_g$ as $I_g=\frac{1}{8\pi^2}\int Tr(Q,\wedge
M_g+\frac{1}{3}Q'\wedge Q'\wedge Q')$ where
$M_g=\Omega_gM\Omega^{-1}$.

Thus we have

\be
I_g-I=\frac{1}{24\pi^2}\int Tr[d\Omega_g \Omega_g^{-1}\wedge
 d\Omega_g \Omega_g^{-1}\wedge d\Omega_g \Omega_g^{-1}] \ee which is
the second Chern Class which describes the winding number of the
manifold. This implies  that the integral of the invariant $tr
M^{\mu\nu}{\cal{M}}^{\mu\nu}$ will lead us to just the one
appropriate Pontraygin invariant for the Yang-Mills gauge group
that is used for the dynamics.  From the minimal coupling
prescription we have given above, we can write:
\begin{widetext}
\be
\int_{M}tr(M_{\mu\nu}{\cal{M}}^{\mu\nu})=\int_{M}tr(F_{\mu\nu}{\cal{F}}^{\mu\nu}
+
\frac{2m}{g}\int_{M}tr(S_{\mu\nu}{\cal{F}}^{\mu\nu})+\frac{m^{2}}{g^{2}}\int_{M}tr(S_{\mu\nu}{\cal{S}}^{\mu\nu})\ee
\end{widetext}
where ${\cal{M, F, S}}$ are the duals of $M, F, S$ respectively and
the integral is over spacetime. Thus in the non-Abelian magneto
fluid we may associate this  non zero Pontryangin index with a
non-Abelian magneto fluid helicity implying the existence of stable
self confining non dissipative solutions. In fact, the non
triviality of the Hopf invariant ensures that flux lines can be
knotted and solitonic configurations are inevitable. These can have
a number of consequences, such as  the existence of glueballs, as
knotted solitons having the non-Abelian helicity as a topological
quantum number, which may survive in the quark gluon plasma in the
interior of a heavy ion collision  or in the early universe.
\section{Conclusion.}
We have given the foundations of a consistent theory of non-Abelian
fluid field system,  in which the flow field and the gauge field are
"unified " in a single minimally coupled gauge flow field. We have
shown that this gives rise to a  quantity which is the fluid field
generalization of the non-Abelian Chern Simons term, and shown that
knotted fluid-field non-Abelian solitons may exist. We can, by using
standard techniques in pure Yang Mills theories,find explicit forms
of these and  look for phenomenological signatures in the context of
QGP. The formalism is simple and unified and should lead to new and
interesting phenomenon such as non-Abelian Alfven waves and other
non-Abelian counterparts of magnetohydrodynamics which may lead to
new signals for collective flow in the QGP. Such studies are under
investigation.

\end{document}